\documentstyle[prl,preprint,aps,epsf]{revtex}
                                  
\begin{document}
 
\title{Electron Spin Decoherence in Bulk and Quantum Well
Zincblende Semiconductors}
\author{Wayne H. Lau, J. T. Olesberg, and Michael E. Flatt\'e}
\address{Department of Physics and Astronomy, University of Iowa, Iowa City, IA 52242}

\maketitle
\begin{abstract}
A theory for longitudinal ($T_1$) and transverse ($T_2$)
electron spin coherence times in 
zincblende semiconductor quantum wells is developed based on a
non-perturbative nanostructure model solved in a
fourteen-band restricted basis set.
Distinctly different dependences of coherence times on
mobility, quantization energy, and temperature are found
from previous calculations.
Quantitative agreement between our calculations and
measurements is found for GaAs/AlGaAs, 
InGaAs/InP, and GaSb/AlSb quantum wells. 
\end{abstract}

\vfill\eject

\tightenlines

The recent observation of long-lived ($>100$~ns) spatially
extended ($>100$~$\mu$m) coherent spin states in
semiconductors suggests the possibility of manipulating
nonequilibrium electron coherence to an unprecedented degree
in a solid.\cite{background,kikkawa97,kikkawa98,kikkawa99}
These spin states interact 
with light, and thus can be used to generate a
host of novel dynamic nonlinear optical and 
electrical effects.\cite{othereffects}
The magnitude and persistence
of such effects is governed partly by the spin
coherence times $T_1$ and $T_2$, describing the decay of
longitudinal and transverse spin order, respectively.
Ultrafast optical measurements have been performed of both 
$T_1$ and $T_2$, although in different geometries
\cite{kikkawa97,kikkawa98,kikkawa99,Wagner,Miller,terauchi99,tackeuchi,vandriel}.

To guide further efforts in the controllable manipulation of
room-temperature spin coherence it is essential to have a
{\it quantitative} theory of spin decoherence. 
Direct quantitative comparison of the current
theory with experiment has been rare for quantum wells, for an 
independent measurement of the mobility of the quantum well is required.
Recently such a comparison was made for room-temperature
electron spin lifetimes in {\it n}-doped GaAs/AlGaAs 
multiple quantum wells (MQWs)\cite{terauchi99}. In addition to measured
$T_1$'s one order of magnitude longer than those predicted from current
theory there were discrepancies in the power law dependences of $T_1$ 
on mobility and confinement energy.  

This Letter provides
the desired quantitatively accurate theory of spin
decoherence for quantum wells and clarifies the relationship
between $T_1$ and $T_2$ in these systems.  
Our results are in
excellent agreement with experimental measurements on GaAs
quantum wells\cite{terauchi99}, not only in the previously unexplained general
trends, but also in the absolute magnitude. We also find excellent
agreement with measurements on InGaAs/InP\cite{tackeuchi}
 and GaSb/AlSb\cite{vandriel}
quantum wells, whereas previous calculations disagree by an
order of magnitude. Finally we find unexpected
trends in the spin coherence times with temperature which
may explain other puzzling experimental results.

The mechanism of electron spin decoherence we consider
occurs via the spin precession of carriers with finite crystal
momentum {\bf k} in the effective {\bf k}-dependent crystal 
magnetic field of an 
inversion-asymmetric material. 
A signature of this mechanism is that in the  ``motional
narrowing'' regime, where orbital scattering times $\tau$
greatly exceed spin decoherence times, $T_1\propto \tau^{-1}$.
Thus cleaner samples have shorter spin coherence times. In
III-V bulk\cite{Optical} and quantum well structures\cite{terauchi99}
nanostructures this trend has been observed in samples of
varying mobility at and near room temperature.

D'yakonov and Perel' have developed a theory for $T_1$
based on this mechanism for bulk
zincblende semiconductors, assuming 
orbital coherence is lost after each scattering event,
and assuming a thermal
distribution of electrons\cite{DP}. This work was later extended
(with further approximations)
to quantum wells\cite{dyakonov86} by D'yakonov and
Kachorovskii (DK). Thus the two categories of approximation
in DK theory are (1) the method of handling the orbital
degrees of freedom and (2) the quantum well electronic structure.
For example, if some {\it orbital} coherence or nonthermal
occupation were maintained
after each scattering event, then because the electron's
orbital degrees of freedom are
entangled with its spin, only a nonequilibrium
calculation of orbital degrees of freedom (e.g. Monte Carlo) would produce
accurate results. We find, however, that the sources of error are the 
approximations pertaining to quantum well
electronic structure, and thus
simpler, {\it quantitatively accurate}
calculations of spin coherence may be performed.

Our theory begins 
with the assumption of motional narrowing.
In the motional narrowing regime the electronic spin system is subject to
an effective {\it time-dependent,} randomly oriented magnetic field
{\bf H} which changes direction with a time $\tau$ that is
much shorter than the precession time of either the
constant applied field ${\bf H_o}$ or the random field.
The coherence times depend on the transverse ($H_\perp$) and
longitudinal ($H_\parallel$)
components of the random field, according to \cite{Yafet}
\begin{eqnarray}
T_1^{-1} &\propto& (H_\perp^2)\tau,\label{t1}\\
T_2^{-1} &\propto& ([H_\perp^2]/2 + H_\parallel^2)\tau,\label{t2}
\end{eqnarray}
where the constant of proportionality is the same for
Eqs.~(\ref{t1}) and (\ref{t2}). 
In a crystal with
inversion asymmetry and spin-orbit coupling there is a spin
splitting described by the Hamiltonian $H=\hbar{\bf
\Omega}({\bf k})\cdot {\bf \sigma}/2$, where 
${\bf\Omega}({\bf k})$ is a
{\it momentum-dependent} effective
magnetic field.  As the electron
is scattered from ${\bf k}$ to ${\bf k'}$ via ordinary
orbital (not spin-dependent) elastic scattering, the effective
magnetic field changes direction with time. 
If the crystal is cubic, then $H_x^2=H_y^2=H_z^2$, so
$H_\perp^2 = 2H_\parallel^2$ and $T_2=T_1$. 
The relationship between $T_1$ and $T_2$ differs, however,
for systems of lower symmetry, such as quantum wells.  For a (001)
grown quantum well the fluctuating field along the
growth direction vanishes, and
\begin{eqnarray}
T_1^{-1}(\alpha) &=& T_{1}^{-1}(\alpha=0)(1+\cos^2\alpha)/2,\\
T_2^{-1}(\alpha) &=& T_{1}^{-1}(\alpha=0)(2+\sin^2\alpha)/4,
\end{eqnarray}
where $\alpha$ is the direction between ${\bf H_o}$ and the
growth direction. Thus $T_2$ ranges from $2T_1/3$ to $2T_1$
depending on $\alpha$.  In contrast, for (110) grown quantum
wells the effective crystal magnetic field is entirely along the growth
direction, and
\begin{eqnarray}
T_1^{-1}(\alpha) &=& T_{2}^{-1}(\alpha=0)\sin^2\alpha,\\
T_2^{-1}(\alpha) &=& T_{2}^{-1}(\alpha=0)(1+\cos^2\alpha)/2.
\end{eqnarray}
Thus although $T_1^{-1}(\alpha=0)$ vanishes, the same is
not the case for $T_2^{-1}$ for any $\alpha$.

Calculations of $T_1$ require knowledge of both
$\tau$ and ${\bf \Omega}({\bf k})$. 
The effective time
for field reversal 
($\tau_\ell$)
depends on the angular index $\ell$
of the field component ($\Omega_\ell$). For example, an $\ell=1$
component ($\Omega_1$)
requires a $180^{\rm o}$ 
change in the angle of ${\bf k}$ to change sign, whereas an $\ell=3$
component ($\Omega_3$) only requires a $60^{\rm o}$ change,
so typically $\tau_3<\tau_1$. Thus
\begin{equation}
{1\over T_1} = {1\over n}\int 
D(E)f(E)[1-f(E)]
\sum_{\ell}
\tau_\ell(E)\Omega_\ell^2(E)dE,
\label{t1qw}
\end{equation}
where $f(E)$ is the Fermi occupation function, $D(E)$ is the
density of states, $n$ is the
electron density, and 
the scattering rates
$\tau_\ell^{-1}(E) = \int_{-1}^1
\sigma(\theta,E)(1-P_\ell(\cos\theta))d\cos\theta$ for bulk
semiconductors and 
$\tau_\ell^{-1}(E) = \int_{0}^{2\pi}
\sigma(\theta,E)(1-\cos[\ell\theta])d\theta$ for (001) quantum
wells. 
For both bulk and quantum wells 
the functional form of the scattering cross-section
$\sigma(\theta,E)$
is taken from standard expressions for
ionized impurity (II), neutral impurity (NI --- such as
arises from quantum well interface roughness), or optical phonon (OP)
scattering. 
The $\tau_{\ell}$'s differ for different
mechanisms (e.g., for a quantum well
$\tau_1/\tau_\ell = \ell^2$ for II,
$\tau_1/\tau_\ell = \ell$ for OP,
and $\tau_1/\tau_\ell = 1$ for NI scattering).
The magnitude of $\sigma(\theta,E)$ is obtained from
the mobility, 
\begin{equation}
\mu = (e/mn)\int D(E) f(E)[1-f(E)] \tau_1(E)E dE.\label{mobility}
\end{equation}

We obtain ${\Omega}_\ell(E)$
from a non-perturbative calculation in a 
fourteen-band basis\cite{dissert}.
This basis, which is the minimum required to
generate spin splitting nonperturbatively, 
consists of two conduction antibonding $s$ states $(\overline{s})$, 
six valence (bonding) $p$ states, and six antibonding $p$ 
states $(\overline{p})$. Such a basis has, for example, been
used to analyze spin-splitting in
heterostructures\cite{Ross}.  The Hamiltonian 
is well-known, and can be found in Refs.~\cite{fourteen,dissert}.
The parameters that enter this Hamiltonian include the
zone-center energies of the constituent bulk semiconductors
and the momentum matrix elements between bands, which 
are obtained from the conduction band mass, the heavy-hole
mass, and the $g$-factor. 
Time reversal invariance requires
${\bf \Omega}({\bf k}) = - {\bf\Omega}(-{\bf k})$, so
$\Omega_\ell = 0$ for even $\ell$.

For quantum wells
the electronic structure is obtained 
by expressing the electronic states as spatially-dependent
linear combinations of the fourteen states in the basis. 
The full Hamiltonian is projected onto this restricted
basis set, which produces a set of fourteen coupled differential
equations for the spatially-dependent coefficients of
the basis states (generalized envelope functions). These equations are
then solved in Fourier space in a similar method to that of
Winkler and R\"ossler.\cite{winkler} Further details are
available elsewhere\cite{dissert}.

\begin{figure}[h]
\centerline{\epsfxsize=5.2in\epsffile[100 150 530 656]{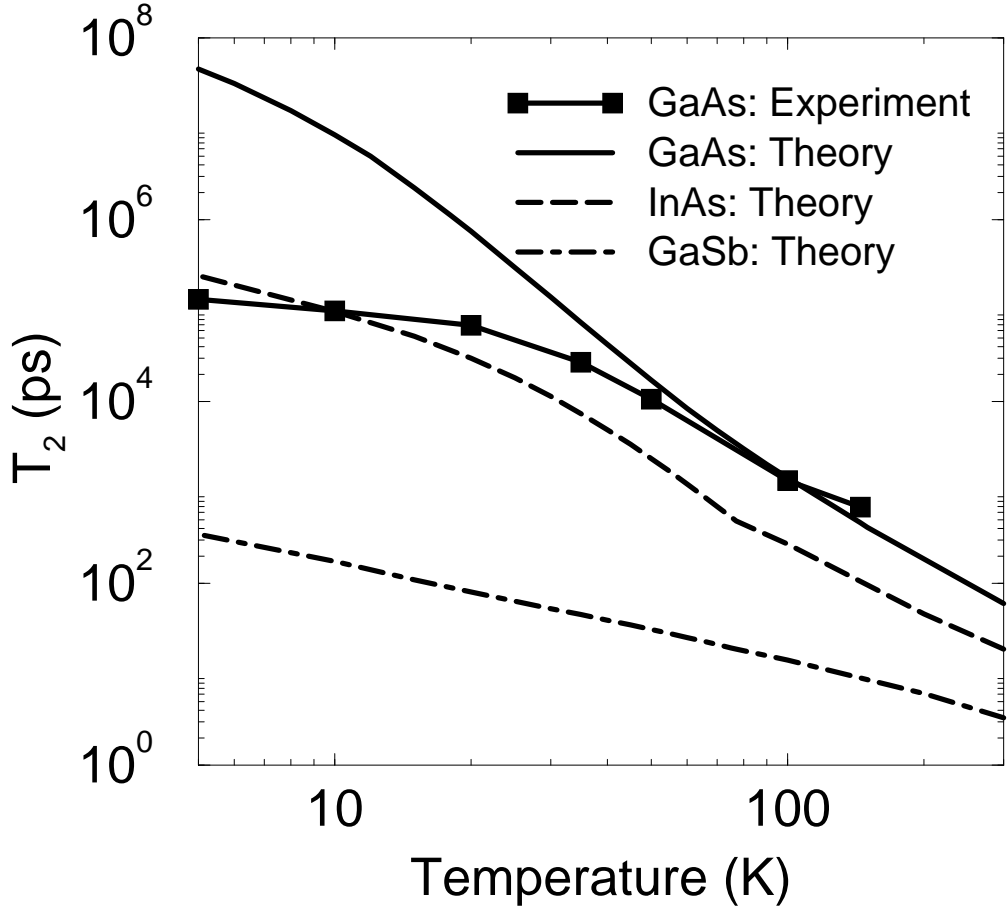}}
\caption[]{$T_2$ in bulk III-V
semiconductors as a function of temperature.  
Solid with squares and solid
lines respectively represent the results of experiments
\cite{kikkawa98} and the non-perturbative theory for bulk GaAs
at the electron density $n=1.0 \times 10^{16}$ cm$^{-3}$. 
Also shown are results for 
bulk InAs at $n=1.7 \times 10^{16}$ cm$^{-3}$ and bulk GaSb
at $n=1.49 \times 10^{18}$ cm$^{-3}$, which are indicated 
with dashed and dot-dashed 
lines respectively.
The difference in slope between GaSb and GaAs occurs because
GaSb is degenerate for this density.
The tabulated mobilities\cite{madelung96} 
for InAs and GaSb extend only to 77K, so at lower temperatures
$\tau_3(E)$ was assumed to have the same value as at 77K.
}
\label{bulkZB}
\end{figure}

For bulk semiconductors the relevant electronic states for
spin decoherence are near the bulk band
edge, and thus perturbative expressions for $\Omega_\ell^2(E)$ for
these bulk semiconductors ($\Omega_1^2(E) = 0$,
$\Omega_3^2(E)\propto E^3$)\cite{DP} are identical to 
those obtained from a full fourteen-band calculation within
numerical accuracy.  Shown in
Fig.~\ref{bulkZB} are calculated $T_2$'s
for GaAs, InAs, and GaSb assuming II scattering. 
The agreement with experimental measurements\cite{kikkawa98}
for GaAs at the higher
temperatures is quite good, whereas for low
temperatures other spin relaxation mechanisms are expected
to dominate. The smaller $T_2$'s
in InAs and GaSb are due partly to the 
larger conduction spin splitting, which
originates from a larger ratio of the spin-orbit 
coupling $\Delta$ to the band gap
$E_g$  (see Ref.~\cite{cardona88} 
on perturbative expansions of spin splittings).
The agreement between calculated and
measured $T_2$'s in Fig.~\ref{bulkZB} indicates that
the spin splitting of bulk GaAs is well described by our
model. 

We now contrast our results for quantum wells with those of
DK theory.
The DK theory for (001) quantum wells is derived as follows.
First, negligible
penetration of the electronic states into the barriers is
assumed, so
$E_1\ll \Delta E_c$, where $E_1$ is the confinement energy
of the first quantum well state and $\Delta E_c$ is the
conduction band offset. Then the
perturbative expressions\cite{ivchenko}
$\Omega_1^2(E)\propto E(4E_1-E)^2$ and $\Omega_3^2(E)\propto
E^3$ are used. Furthermore the energies of relevant states
are assumed to be $\ll E_1$, and thus
{\it (i)} the contribution from $\Omega_3(E)$ is 
ignored, and {\it (ii)} it is assumed that $\Omega_1^2(E)\propto
E$.  It is not generally recognized that the conditions
$kT\ll E_1\ll \Delta E_c$ are quite restrictive and are
difficult to satisfy at room temperature.

\begin{figure}[h]
\centerline{\epsfxsize=5.6in\epsffile[80 150 520 656]{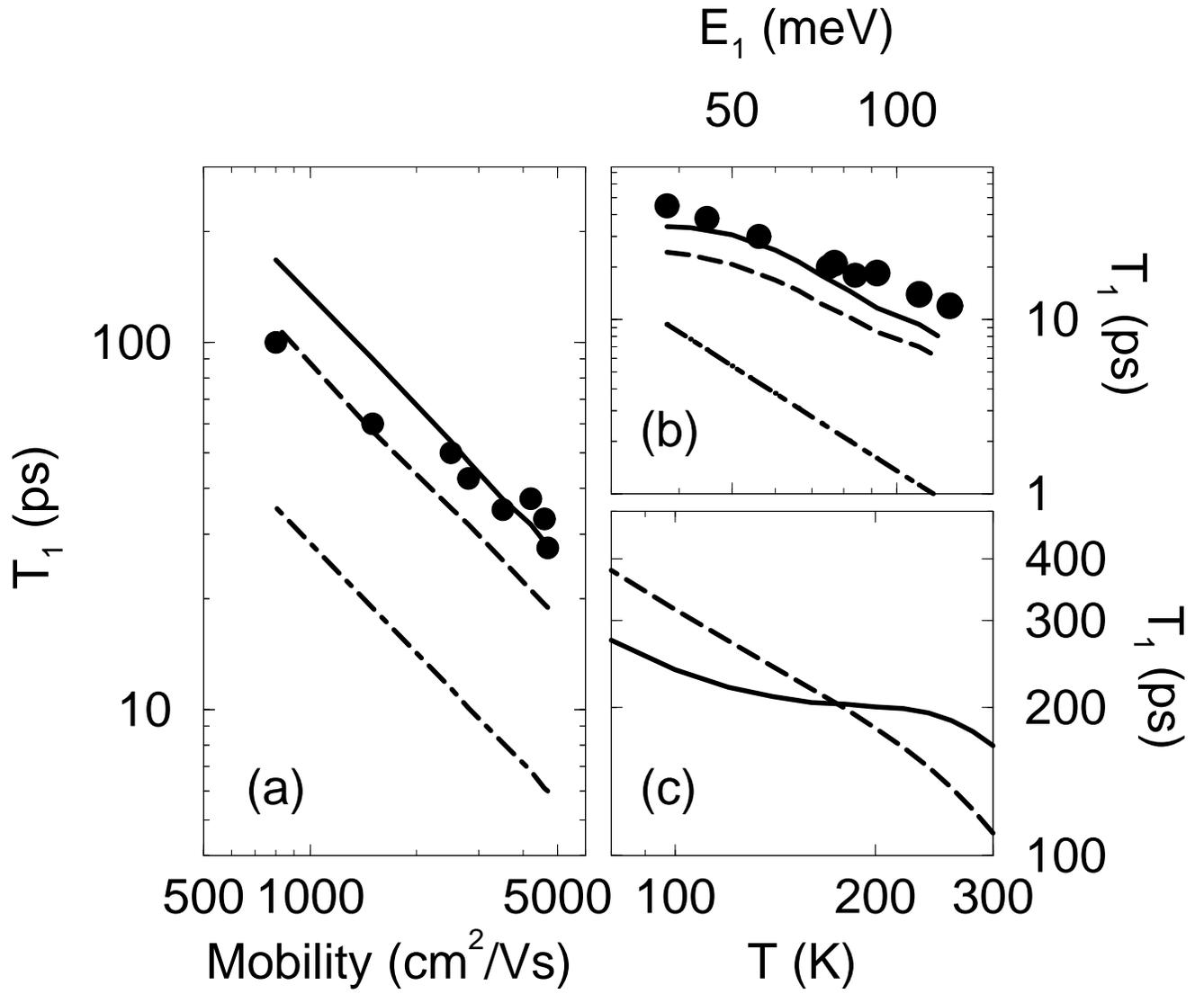}}
\caption[]{$T_1$ as a function of (a) mobility, (b) confinement
energy, and (c) temperature, for 
$75$\AA\ GaAs/Al$_{0.4}$Ga$_{0.6}$As MQWs at room temperature. 
Closed circles represent the results of 
experiments\cite{terauchi99}. The non-perturbative theory
results
with OP scattering (solid lines) and NI scattering (dashed
lines) are shown, as well as the DK theory results
(dot-dashed lines).}
\label{mqwt1}
\end{figure}

The resulting $T_1^{-1}$ [Eq.~(\ref{t1qw})] under the DK
assumptions is proportional to 
the mobility {\it independent of the dominant scattering mechanism}
[see Eq.~(\ref{mobility})]. In addition, $T_1^{-1}$ is
proportional to $E_1^2$.
These trends are not supported by recent
experimental measurements \cite{terauchi99,caveat} on
$75$\AA\ {\it n}-doped GaAs/Al$_{0.4}$Ga$_{0.6}$As MQWs
at 300K [shown in Fig.~\ref{mqwt1}(a,b) (filled circles)]. 
In both cases the experimental trends are weaker than the
predicted theoretical ones.
Calculations are shown in Fig.~\ref{mqwt1}(a,b) based on our more
general theory using OP (solid line)
and NI (dashed line) as the dominant
process determining the mobility. 

Our results agree with experiment if one assumes a shift
from OP to NI scattering as the mobility drops
---  this is the origin of the unusual experimental dependence of $T_1$
on the mobility.  The weaker dependence of $T_1$ on $E_1$ in
our theory versus DK theory in Fig.~\ref{mqwt1}(b) is due to wavefunction
penetration into the barriers and non-perturbative effects.
We emphasize as well the key role of temperature studies of
the mobility in analyzing the temperature dependence of spin
coherence. In Fig.~\ref{mqwt1}(c) the calculated $T_1$ for one sample with
a given room-temperature mobility is
presented as a function of temperature for NI and OP
scattering. In particular the OP results appear relatively insensitive
to temperature from 100-250K --- this is due to the rapid decrease in the
mobility from OP scattering with increasing temperature. This
may play a role in the weak temperature dependence seen in
Ref.~\cite{kikkawa97}.

\begin{figure}[h]
\centerline{\epsfxsize=5.5in\epsffile[60 150 530 656]{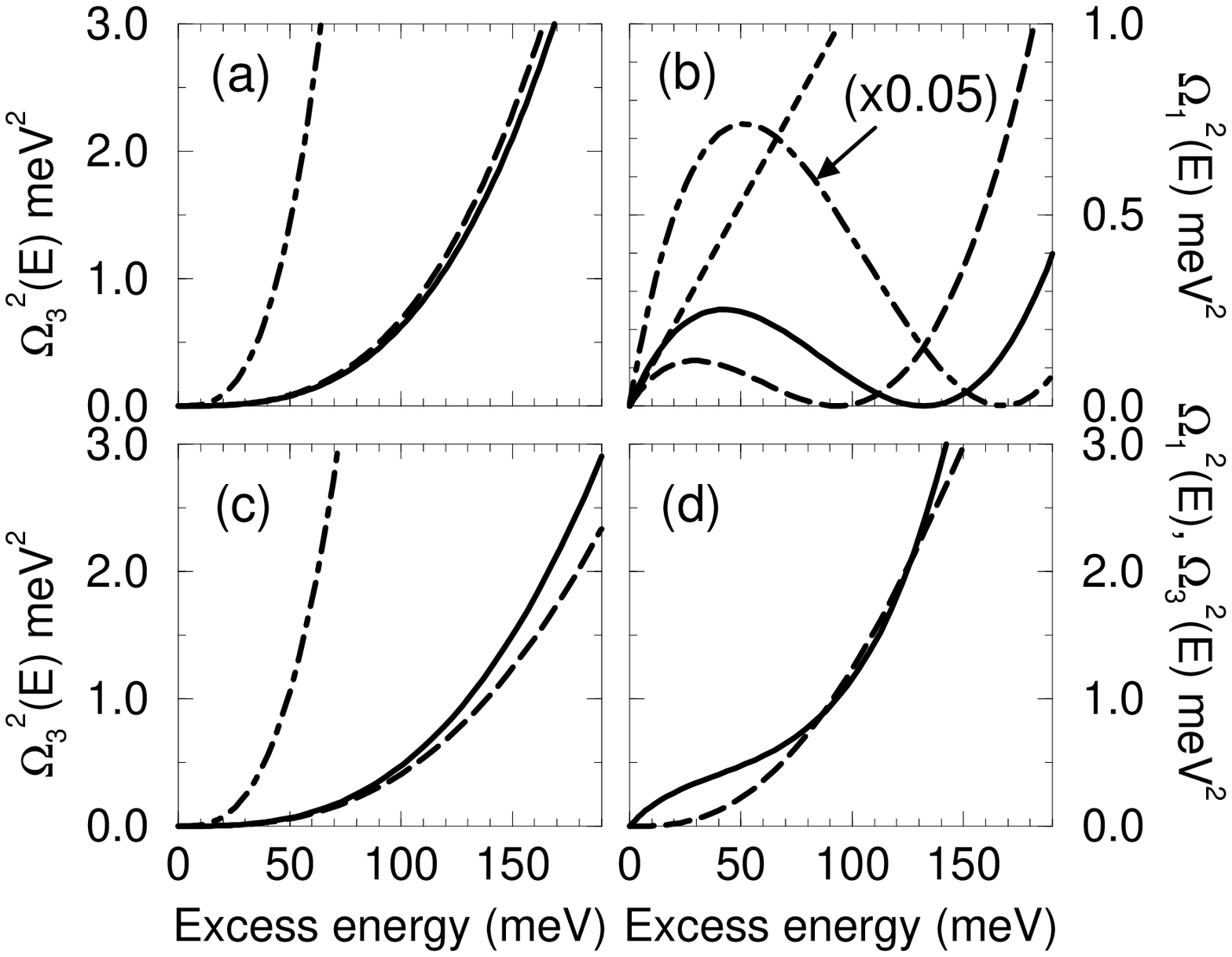}}
\caption[]{$\Omega_1^2(E)$ and $\Omega_3^2(E)$ for several
structures. (a) $\Omega_3^2(E)$ for bulk GaAs (solid), InAs
(dashed), and GaSb (dot-dashed). (b) $\Omega_1^2(E)$ 
for GaAs (solid), InGaAs (long dashed), and GaSb (dot-dashed)
quantum wells described in Table~\ref{tab}. The short-dashed line
is the DK approximation for the GaAs quantum well. 
(c) $\Omega_3^2(E)$ for the same
three structures as (b). (d) $\Omega_1^2(E)$ (solid) and
$\Omega_3^2(E)$ (dashed) for a thin-layer InAs/GaSb superlattice.
\label{omegas}}
\end{figure}

Figure~\ref{omegas} compares the energy dependence of
$\Omega_1^2(E)$ and $\Omega_3^2(E)$ for several additional material
systems.
The cubic dependence of $\Omega_3^2(E)$ for the three bulk
semiconductors is confirmed in Fig.~\ref{omegas}(a).
Fig.~\ref{omegas}(b), however, shows that for quantum
wells $\Omega_1^2(E)$ is
only linear (short dashed line for the GaAs MQW) for a small energy 
range ($\sim 20$~meV) above the band edge
before it begins to deviate. 
More energetic states than this certainly contribute to the spin
coherence times at room temperature. 
The wider the well the lower the energy where $\Omega_1^2(E)$ 
deviates from linear behavior, as it approaches a bulk-like
$E^3$ behavior.  
$\Omega_3^2(E)$ 
for these structures is shown in Fig.~\ref{omegas}(c), and
is comparable in magnitude to $\Omega_1^2(E)$.
As the wells become narrower, even the perturbative expressions
for $\Omega_3$ and $\Omega_1$ break down. 
Figure~\ref{omegas}(d) shows $\Omega_1^2(E)$ and
$\Omega_3^2(E)$ for a thin-layer InAs/GaSb superlattice,
indicating very different behavior from the other
structures, poorly reproduced by even the general forms of
the perturbative expression.

Table~\ref{spintimes} presents calculations and experimental
measurements of $T_1$ for these material systems. The
order of magnitude discrepancy between DK calculations and
measurements occurs here as well. 
Given the uncertainty in experimental mobilities and
densities, the agreement of our calculations with experiment for 
both NI and OP
scattering is good for 
all systems, and is much better than DK theory.
Note that OP and NI scattering calculations in the full
theory differ 
from each other by factors of of up
to 2 (due to differences in $\tau_\ell(E)$), whereas 
all scattering mechanisms produce the same
result in DK theory. 
As expected,
for several systems the $T_1$'s are much shorter at higher
electron densities, for as the carrier distribution is
spread further from zone center the effective crystal magnetic
fields increase.
The DK approximation {\it (i)} can be
evaluated by comparing OP$_1$ to OP and NI$_1$ to NI, where
calculations using all terms up to $\ell$ are
designated OP$_\ell$ and NI$_\ell$.
The difference is up to 40\%. Approximation
{\it (ii)}, however, produces a discrepancy between the DK result and
both NI$_1$ and OP$_1$
which can greatly exceed an order of magnitude.

These calculations consider decoherence arising from the bulk
inversion asymmetry (BIA) of the constituent materials. We
have considered symmetric wells, so another source of
inversion asymmetry, the
structural inversion asymmetry (SIA), does not play a role. 
In other structures, such as
single-interface heterostructures, SIA may dominate\cite{PW}.
Interface bonding asymmetry (native
interface asymmetry, or NIA), which arises
in non-common-atom structures, could play a
role in systems II, IV, or V. 
The NIA spin splitting for perfect interfaces
(imperfect interfaces reduce the NIA contribution)
has been calculated for System II  
in Ref.~\cite{vervoort99}. 
By comparing with Ref.~\cite{vervoort99} we
find the spin splitting of this quantum well is dominated by BIA.

We have presented a {\it quantitatively accurate}
non-perturbative nanostructure theory for
electron spin relaxation in bulk and quantum well zincblende
semiconductors based on a fourteen band model.
The calculated electron spin lifetimes in 
III-V semiconductor bulk and quantum well materials are in agreement 
with experimental 
measurements, indicating the importance of accurate 
band structure calculations for 
zincblende type nanostructures. 

We would like to acknowledge discussions with D. D.
Awschalom, T. F. Boggess, J.~M.~Byers, J. M. Kikkawa,
and A. Smirl. We would also like to thank D. D. Awschalom
and J.~M.~Kikkawa for providing mobility data for the bulk GaAs
sample of Ref.~\cite{kikkawa98}.
This work was supported in part by the Office of 
Naval Research through Grant No. N00014-99-1-0379.

\begin{table}
\caption[]{Spin coherence times $T_1$ (ps) for several
structures, I: a 75\AA\ GaAs/Al$_{0.4}$Ga$_{0.6}$As 
MQW\cite{terauchi99}, II: a 70\AA\ 
In$_{0.53}$Ga$_{0.47}$As/97\AA\ InP MQW\cite{tackeuchi},
III: an 80\AA\ GaSb/80\AA\ AlSb MQW\cite{vandriel}, 
IV: a 51\AA\ GaAs$_{0.19}$Sb$_{0.81}$/80\AA\
AlSb MQW\cite{vandriel},
and V: a 21.2\AA\ InAs/36.6\AA\ GaSb
superlattice.
Calculated times are shown for a given total electron density
(n.d. indicates nondegenerate) using DK theory
(DK), and the nonperturbative theory with
optical phonon (OP) and neutral impurity (NI)
scattering. The subscript $\ell$ indicates that
only terms up to $\Omega_\ell$ were used in the
calculation.}
\label{spintimes}
\begin{tabular}{lccccccccc}
&System&Density (cm$^{-3}$)&$\mu$~(cm$^2$/Vs)&Exp.&DK&OP$_1$&OP&NI$_1$&NI\\
I&GaAs/AlGaAs&$2.7\times 10^{17}$&800&100&27&151&120&162&111\\
II&InGaAs/InP&n.d.&6700&---&1.45&53&37&52&32\\
&&$3.0\times 10^{18}$&6700&2.6&0.21&6.0&4.9&6.9&4.0\\
III&GaSb/AlSb&n.d.&2000&---&0.59&1.9&1.8&1.5&1.4\\
&&$2.8\times 10^{18}$&2000&0.52&0.09&0.64&0.55&0.88&0.53\\
IV&GaAsSb/AlSb&n.d.&2000&---&0.09&0.53&0.52&0.44&0.43\\
&&$3.4\times 10^{18}$&2000&0.42&0.01&1.9&1.4&1.7&0.87\\
V&InAs/GaSb&n.d.&3000&---&0.38&0.77&0.77&1.7&1.6\\
\end{tabular}
\label{tab}
\end{table}

\begin{references}
\bibitem{caveat}The experimental results have been
adjusted from Ref.~\cite{terauchi99}, for the authors
defined an effective spin
flip time for a single spin, $\tau_s=2T_1$, and plotted their results for
$\tau_s$. 
The DK calculation (dot-dashed line) is of $T_1$,
so in addition to the errors in trends, the discrepancy in
the magnitude of $T_1$ (using our values of the
confinement energy) is about a factor of 4 in Fig.~\ref{mqwt1}(a,b).
\end{references}
\end{document}